\documentclass[aps,floatfix,footinbib,longbibliography,twocolumn,superscriptaddress, preprintnumbers]{revtex4-1}

\usepackage{dsfont}
\usepackage{bbm}
\usepackage{amsmath}
\usepackage{amsfonts}
\usepackage{amssymb}
\usepackage{graphicx}
\usepackage{color}
\usepackage{accents}
\usepackage[colorlinks=true, citecolor=blue]{hyperref}
\usepackage[all]{hypcap}
\usepackage{subfigure}
\usepackage[normalem]{ulem}

\newcommand{\Tr}{\mbox{Tr}}

\begin{document}

%\preprint{APS/123-QED}
%\title{On Machine Learning for Quantum Evolution}
\title{Data-Driven Time Propagation of Quantum Systems with Neural Networks}
\author{James Nelson}\email{janelson@tcd.ie}
\affiliation{School of Physics, AMBER and CRANN Institute, Trinity College, 
	Dublin 2, Ireland}
\author{Luuk Coopmans}\email{coopmanl@tcd.ie}
\affiliation{School of Physics, AMBER and CRANN Institute, Trinity College, 
	Dublin 2, Ireland}
\affiliation{Dublin Institute for Advanced Studies, School of Theoretical Physics, 10 Burlington Rd, Dublin, Ireland}
\author{Graham Kells}\email{gkells@stp.dias.ie}
\affiliation{Dublin Institute for Advanced Studies, School of Theoretical Physics, 10 Burlington Rd, Dublin, Ireland}
\affiliation{Dublin City University, School of Physical Sciences, Glasnevin, Dublin 9, Ireland}
\author{Stefano Sanvito}\email{sanvitos@tcd.ie}
\affiliation{School of Physics, AMBER and CRANN Institute, Trinity College, 
	Dublin 2, Ireland}

\date{\today}

\begin{abstract}
We investigate the potential of supervised machine learning to propagate a quantum system in time. While Markovian 
dynamics can be learned easily, given a sufficient amount of data, non-Markovian systems are non-trivial and their 
description requires the memory knowledge of past states. Here we analyse the feature of such memory by taking a 
simple 1D Heisenberg model as many-body Hamiltonian, and construct a non-Markovian description by representing 
the system over the single-particle reduced density matrix. The number of past states required for this representation 
to reproduce the time-dependent dynamics is found to grow exponentially with the number of spins and with the density
of the system spectrum. Most importantly, we demonstrate that neural networks can work as time propagators at any time
in the future and that they can be concatenated in time forming an autoregression. Such neural-network autoregression 
can be used to generate long-time and arbitrary dense time trajectories. Finally, we investigate the time resolution needed 
to represent the system memory. We find two regimes: for fine memory samplings the memory needed remains constant, 
while longer memories are required for coarse samplings, although the total number of time steps remains constant. The 
boundary between these two regimes is set by the period corresponding to the highest frequency in the system spectrum, 
demonstrating that neural network can overcome the limitation set by the Shannon-Nyquist sampling theorem.
\end{abstract}

\maketitle

\section{Introduction}

The notion of a memory kernel arises frequently in the study of dynamical models, where one attempts to 
understand or recreate dynamical behaviours using only a subset of the total available degrees of freedom. 
In particular, a kernel becomes necessary in situations where the subset of information available at a given 
instant in time is not enough to predict the subsequent behaviour. In open quantum systems, where a restricted
number of degrees of freedom is in contact with a bath, one formally understands this subset restriction as the 
tracing out of the bath. Here, the need of a memory kernel to obtain accurate long-time trajectories is a defining 
criteria that distinguishes Markovian from non-Markovian dynamics. 

A number of key techniques and approaches have emerged enabling a systematic study of this effect. One such 
family are the so-called projection techniques (e.g  the Nakajima-Zwanzig equation \cite{Nakajima1958,Zwanzig1960,Prigogine1962} 
and their time-convolutionless variants \cite{Shibata1977,Shibata1980}) that render the history (memory) within an 
integro-differential equation and can be used to provide a perturbative justification for the Markovian behaviour.
Another approach for exploring memory effects is to embed the system into an enlarged Hilbert space such that the 
non-Markovian dynamics of the reduced system can be represented via the Markovian evolution of an enlarged whole 
\cite{Ciccarello2013,McCloskey2014, Kretschmer2016, Lorenzo2017, Cakmak2017,Campbell2018}. In this respect, a 
useful class of embeddings are the so-called collision models, where ancillas representing the bath interact in a 
time-dependent fashion. Here the concept of memory depth naturally arises, when one allows these ancillas to interact 
between themselves.

Both the projection and embedding schemes represent intuitive and inherently human-learned/reasoned approaches for 
investigating the interplay between a quantum system and its associated bath. Given the now widespread application of 
machine learning (ML) in the domain of classical and quantum dynamics, it is natural that one might apply machine learning 
to the same domain. For a selective snapshot of this field we refer the reader to some representative published literature \cite{mardt2018vampnets,breen2020newton,carleo2017solving,hartmann2019neural,lin2021scaling,banchi2018modelling,Luchnikov2020} 
and references therein. 

When looking at the non-Markovianity of a time evolution, in particular with embeddings and memory kernels, an approach 
that has recently emerged is to construct a fixed neural network (NN) that learns the behaviours of local observables and 
then attempts to match the time dependence of the observables beyond the training time. For instance, in reference 
\cite{Mazza2021} this data-driven method was applied to learn time-local generators of the dynamics of open quantum 
systems. Similarly, in \cite{Luchnikov2021} a related approach was used to find an effective Markovian embedding from 
which several important properties and the exact dynamics of the reduced system could be extracted.

One of the key advantages of these data-driven methods is their universality. In fact, machine-learning time propagators 
can be used with any method to generate quantum evolution and show promise for use in investigations into the nature 
of the non-Markovian evolution. In this respect, the methodology blends features of Markovian embeddings and 
integro-differential memory kernels, but where the functional that encodes the dynamics is fixed so that the embeddings 
and the kernel/memory encoding are time or state independent. 

In this paper we take a broadly similar approach in that we train a NN on a dataset of time-dependent quantum trajectories 
and then ask it to evolve an arbitrary starting state up to times longer than the final time of the learned trajectories. 
Distinguishing our work from previous studies, we look at the complete collection of reduced single-particle density matrices 
that together make up the full (strongly interacting) system and focus, in particular, on the accuracy of the method with respect 
to the depth of the sampling history that is used as input for the NN. While in general, as we show below, the Markovian dynamics 
of the full system can be learned by only one history sample, for the representation in terms of reduced density matrices this 
is no longer the case. In fact, we find that the sampling history needs to be increased exponentially with system size to keep 
the error rates on the predicted dynamics below a fixed threshold. Aside from studying such memory effects, an advantage 
of this approach is that its malleability allows for non-fixed time sampling and naturally protects against oversampling error. 
Moreover, as the error growth on a given NN is random we can use the idea of neural-network ensembling to monitor and 
reduce errors.

We structure our paper as follows: in section \ref{sec:background} we give an overview of various methods to study memory 
effects in quantum dynamics and introduce our specific setup and method to generate the data of the quantum evolution. 
In section \ref{sec:MK} and \ref{sec:nMK} we present our results for Markovian and non-Markovian dynamics, respectively,
discussing the interplay between the non-locality in time and space. Then we conclude.

\section{Methodology and background}
\label{sec:background}

\subsection{Brief review of existing methods} 

\begin{figure}
	\centering
	\includegraphics[width=0.9\linewidth]{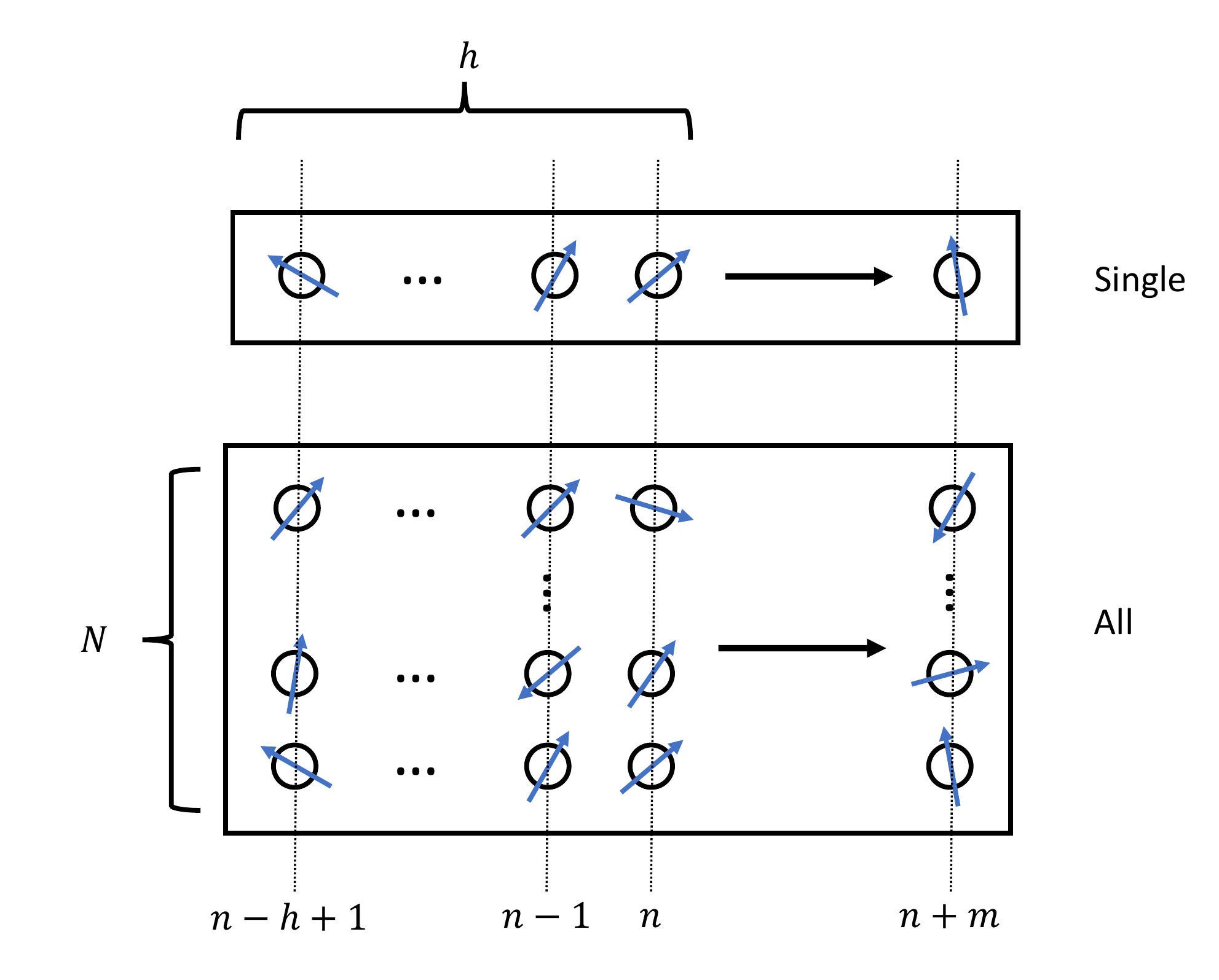}
	\caption{(Color on line) Here the two types of non-Markovian mappings are shown. The small blue arrows symbolise 
	the Block vectors of the individual qubits, while the dashed lines are the time steps. $N$ is the number of qubits, $h$ is 
	the history and $m$ is the future time where we map the evolution to. In the `single' case, the history of a single qubit 
	(evolving in the presence of the others, which then define the {\it environment}) is used to predict its future dynamics, 
	while in the `all' case, the history of all the qubits, is used to predict the future dynamics of the entire system.}
	\label{fig:memory_diagram}
\end{figure}

We consider a quantum system described by its density matrix, $\hat{\rho}$, that consists of a total of $N$ qubits. 
The reduced density matrices of such system can be obtained by tracing out selected degrees of freedom, here 
collectively called the {\it environment}, $\hat{\rho}_\text{red}=\text{Tr}_\text{E}\hat{\rho}$, where `E' stands for some 
environment (see Figure \ref{fig:memory_diagram}). When the traced-out region is all but one qubit we use the 
parameterisation $\hat{\rho}_\text{red}= \frac{1}{2}(\mathbbm{1}+\textbf{r}\cdot \pmb{\sigma})$, where $\textbf{r}$ 
is the Bloch vector. Since the reduced density matrices leave out much information they alone cannot determine the 
future evolution of the entire system. Thus, for a typical dynamical reconstruction of the reduced quantities over time 
one may also need some historical knowledge of past states. Several analytical and numerical techniques have been 
developed to study the memory effects in this type of setup. 

{\em Projection methods:} 
A technique often mentioned in literature to render the past knowledge explicit in the time-evolution equations 
is the Nakajima-Zwanzig projector method \cite{Nakajima1958,Zwanzig1960,Prigogine1962} (see appendix 
\ref{app:NZ} for a quick derivation). When the time-evolution starts ($t=0$) from an initially unentangled system 
and bath, $\hat{\rho}(t=0)=\hat{\rho}_{\text{red}}\otimes\hat{\rho}_{\text{E}}$, this approach allows one to reduce 
the Liouville-Von Neumann equation for the entire system,
\begin{equation}
\label{eq:Liouville_main}
\partial_t \hat{\rho} = \frac{i}{\hbar} \left[ \hat{\rho},\hat{H} \right] = \hat{L} \hat{\rho}\:,
\end{equation}
to an equation containing just the relevant reduced density operator, $\hat{\rho}_{\text{red}}$,
\begin{equation}
\label{eq:NZmain}
\partial_t [\hat{\rho}_{\text{red}}] =  \mathcal{\hat{P}} \hat{L}  [\hat{\rho}_{\text{rel}}] +\int_0^t dt' \mathcal{\hat{K}}(t') [ \mathcal{\hat{P}}\hat{\rho}_{\text{red}}(t-t')]\:,
\end{equation}
where 
\begin{equation}
\mathcal{\hat{K}} (t) = \mathcal{\hat{P}}\hat{L} e^{\mathcal{\hat{Q}} \hat{{L} t}}\mathcal{\hat{Q}}\hat{L} \mathcal{\hat{P}}\:.
\end{equation}
Here we have introduced the Liouvillian operator, $\hat{L}$, and the projectors $\mathcal{\hat{P}}$ and 
$\mathcal{\hat{Q}}=\mathcal{I}-\mathcal{P}$ ($\mathcal{I}$ is the identity), which are defined in terms of 
the reduced density matrix $\hat{\rho}_{\text{red}}$ and a fixed reference state $\hat{\rho}_\mathrm{B}$ 
of the environment degrees of freedom,
\begin{equation}
\hat{\rho}_{\text{red}}= \hat{\mathcal{P}} \hat{\rho} = \text{Tr}_\mathrm{B} ( \hat{\rho} ) \otimes \hat{\rho}_\mathrm{B}\:.
\end{equation}

The memory kernel $\mathcal{\hat{K}}(t)$ includes the effects of the history of the reduced state. Solving the 
Nakajima-Zwanzig equation and finding the memory kernel is in general a computationally hard task. Transfer-tensor 
methods \cite{Cerrillo2014,Gelzinis2017,Gherardini2021} have been recently proposed as efficient techniques to 
reconstruct the memory kernel. In these the individual quantum trajectories are used to learn dynamical maps that can be 
converted into the so-called transfer tensors. The transfer tensors are then directly related to the memory kernel 
of the Nakajima-Zwanzig equation \cite{Cerrillo2014}.  

{\em Embeddings and collision models:}  
An alternative general approach to the study of memory consists in collision models and the so-called Markovian embeddings 
[see, for instance, \cite{Ciccarello2013,McCloskey2014, Kretschmer2016, Lorenzo2017, Cakmak2017,Campbell2018}]. The 
general idea is that the system interacts in a controlled way with an environment comprising of ancillas that, in turn, can interact 
amongst themselves in a variety of ways. In this approach, which has been very fruitful for understanding the boundary between 
Markovian and non-Markovian dynamics, one typically employs some spin Hamiltonian, where the interaction is turned on and 
off sequentially in some predetermined fashion. For the purposes of our study we similarly employ a one-dimensional spin 
description of our system, specifically using the Heisenberg model as generator of the dynamics. Thus the Hamiltonian reads,
\begin{equation}\label{eq:H}
	H = -\frac{J}{2}\sum_{j=1}^N \pmb{\sigma}^{(j)} \cdot\pmb{\sigma}^{(j+1)}\:,
\end{equation}
where $\pmb{\sigma}$ is the vector of Pauli matrices for spin 1/2, $N$ is the total number of qubits and $J$ defines the time 
scale of the problem. In particular here we consider finite periodic spin arrangements,
namely rings of spins. In our case, since our motivation is to understand how well the full dynamics can be predicted by using 
a machine-learning (ML) data-driven technique, no piecewise temporal distinctions are made.  

\subsection{Data-driven methods} 

The data driven approach for predictive dynamics used here and, for instance, in references 
\cite{Mazza2021, Luchnikov2021} has some clear similarities to both the projection and the 
collision methods. One defines $\textbf{r}_{i}^{t}$ as the Bloch vector for the $i$-th qubit at the 
time $t$, which is then discretised into $t_n=n\Delta$ with $n$ an integer and some fundamental 
time $\Delta$, and sets
\begin{equation}
R^{t_n}=R^{n} = (\textbf{r}_{1}^{n}, \textbf{r}_{2}^{n}, ..., \textbf{r}_{N}^{n})\:,
\end{equation}
namely, $R^{t_n}$ is the collection of all Bloch vectors at a given time. We consider two types of 
mappings (in the underbraces we show the dimensions): i) `all', where the entire collection of Bloch 
vectors is propagated in time
\begin{equation}
	\underbrace{(R^{n}, R^{n-1}, ..., R^{n-h+1})}_{3\times N\times h} \rightarrow \underbrace{R^{n+m}}_{3 \times N}\:,
\end{equation}
and ii) `single', where the time-evolution of only one Bloch vector is computed,
\begin{equation}
\underbrace{(\textbf{r}^{n}, \textbf{r}^{n-1}, ..., \textbf{r}^{n-h+1})}_{3\times h} \rightarrow \underbrace{\textbf{r}^{n+m}}_{3}\:.
\end{equation}
Here, the history $h$ is the Markov order, which determines the depth of the time memory, and 
$m$ is the distance into the future where the prediction will take place. Unlike when propagating 
the full many-body wave-function, which grows as $2^N$, both mappings scale linearly with the 
number of qubits $N$. A schematic of both these mappings is shown in Figure~\ref{fig:memory_diagram}.  

In this set-up a Markovian evolution corresponds to the $h=1$ case, while the accurate prediction of an 
increasingly non-Markovian dynamics \cite{Campbell2018} can be obtained by enlarging the memory, $h$. 
In lay terms, the ability of a NN to predict the dynamics from only a partial knowledge of the full density 
matrix (the reduced density matrix) is traded off with a non-locality in time. An important feature of this scheme 
is that we can perform autoregression, namely we can feed the predictions at some given time back into the 
model to make predictions at subsequent times. Such ability makes our mappings to be universal time propagators, 
regardless of whether the dynamics is Markovian or not. Notably, as one can train the model to predict at arbitrary 
times in the future (one can train different networks for different $m$'s) both short- and long-time trajectories are 
accessible, with the error being dictated by the accuracy of the specific network.

Although recurrent NNs have been used previously in the literature to model quantum evolution 
\cite{banchi2018modelling, zhang2019predicting}, here we opt for a fully connected NN to learn the propagator of 
the dynamics, since we only deal with inputs of a fixed size. In particular, we use a two-layer network with 64 nodes 
in each hidden layer, and the Exponential Linear Unit (ELU) activation function, which was found to perform best.  
In order to learn the NN parameters we minimize the trace distance between the predicted and true Bloch 
vectors \cite{NoteD}. The networks have been implemented using PyTorch \cite{paszke2019pytorch}. % Metric

\begin{figure}
	\centering
	\includegraphics[width=0.9\columnwidth]{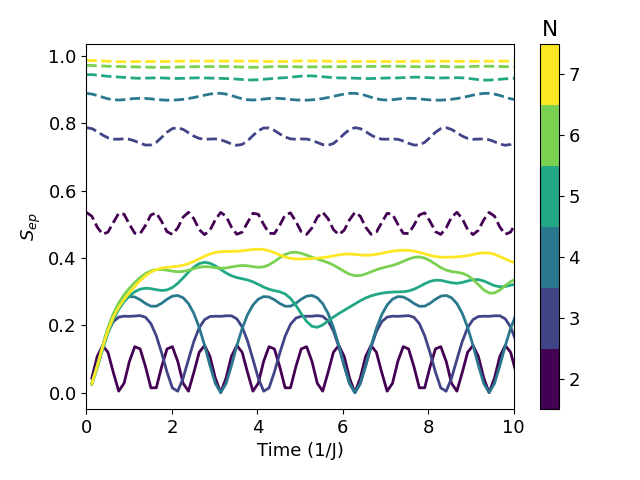}
	\caption{(Color on line) The mean single-particle entropy, $S_\mathrm{sp}=\Tr({\hat{\rho}_{\text{red}}\log{\hat{\rho}_{\text{red}}}})$, 
	as a function of time	for differently initialized states. The dashed lines correspond to randomly generated initial states, while 
	the continuous lines are associated to product initial states. The colour encodes the total number of qubits in the system. 
	Note that product states have vanishing initial entropy, while the entropy of random states grows with the number of qubits 
	considered.}
	\label{fig:sgrowth}
\end{figure}

\subsection{Data generation} 

The time-dependent quantum trajectories used for the training of the NN have been generated by taking two different 
types of pure initial states, $\hat{\rho}(t=0)$, namely random and product states. These are, respectively, constructed 
from initial wavefunctions of the form
\begin{equation}\label{init:rand}
	|\psi(0)\rangle_\mathrm{rand} = \sum_{i_1=0,1} \sum_{i_2=0,1} ... \sum_{i_N=0,1} a_{i_1i_2...i_N}|i_1i_2...i_N\rangle\:,
\end{equation}
\begin{equation}\label{init:prod}
|\psi(0)\rangle_\mathrm{prod} = \sum_{i_1=0,1} a_{i_1}|i_1\rangle \sum_{i_2=0,1} a_{i_2}|i_2\rangle ... \sum_{i_N=0,1} a_{i_N}|i_N\rangle\:,
\end{equation}
where the vector $|i_n\rangle$ spans the Fock space of the $n$-th qubit and the coefficients $a_{i_n}$ and $a_{i_1i_2...i_N}$ 
are selected randomly in both cases. Note that, on the one hand, the single-particle reduced density matrices associated
to product states have all zero Von Neumann entanglement entropy, $\Tr({\hat{\rho}_{\text{red}}\log{\hat{\rho}_{\text{red}}}})$, and 
hence there is no entanglement initially present in the system. On the other hand, random initial states have, on average, a high 
entanglement entropy, as shown in figure~\ref{fig:sgrowth}. 

These initial states are then time evolved with the propagator of the total system (the specific time-dynamic generators 
are discussed below). In particular, at a practical level the equations of motion have been integrated numerically by using 
the {\sc QuTiP} Python package~\cite{QuTiP}. Then, we trace out the environment degrees of freedom (some of the quibits) 
to generate the quantum trajectories of the single qubits. For all our numerical experiments - unless otherwise specified - we 
use a dataset consisting of 11,000 samples for each system size, $N$. These are divided into 8,000 training samples, 
2,000 validation ones, while 1,000 samples are used as test set. We employ the validation set to determine when to halt 
the training. Throughout this work all the results shown are calculated on the test set.

\section{Predicting the evolution of Markovian systems}
\label{sec:MK}

For Markovian dynamics the state of the system at a given time contains sufficient information to predict 
its future evolution. This is the case when the state is described by the wavefunction or by the full density matrix. 
Here, since either the Schr\"odinger or the Von Neumann equation imply a linear mapping between states at different 
times, given a set of trajectories we can use a linear regression to learn the propagator. Let the vector $\textbf{x}_n$ 
represent the wavefunction or density matrix of a system at time step $n$, then the propagator, $P$, is a matrix with 
$\textbf{x}_{n+1}=P\textbf{x}_n$. Given a dataset with $p$ pairs of single time step evolutions 
$\{(\textbf{x}_n^{(1)}, \textbf{x}_{n+1}^{(1)}),...,(\textbf{x}_n^{(p)}, \textbf{x}_{n+1}^{(p)})\}$ we can 
learn the propagator matrix by writing
\begin{equation}\label{lin_reg}
	P = (X_n^TX_n)^{-1}X_n^T X_{n+1}\:,
\end{equation}
where $X_n$ is a matrix, whose $i$-th row is  $\textbf{x}_n^{(i)}$. Thus, we can exactly learn the propagator, if the 
matrix $X_n^TX_n$ is invertible. This is the case when the number of samples, $p$, is greater than the dimension 
of the vector. If $\Omega$ is the dimension of the Hilbert space, then we require $2\Omega-1$ samples to learn the 
wavefunction propagator and $\Omega^2-1$ samples to learn that of the density matrix \cite{Note}.

\begin{figure}[h]
	\centering
	\includegraphics[width=0.9\columnwidth]{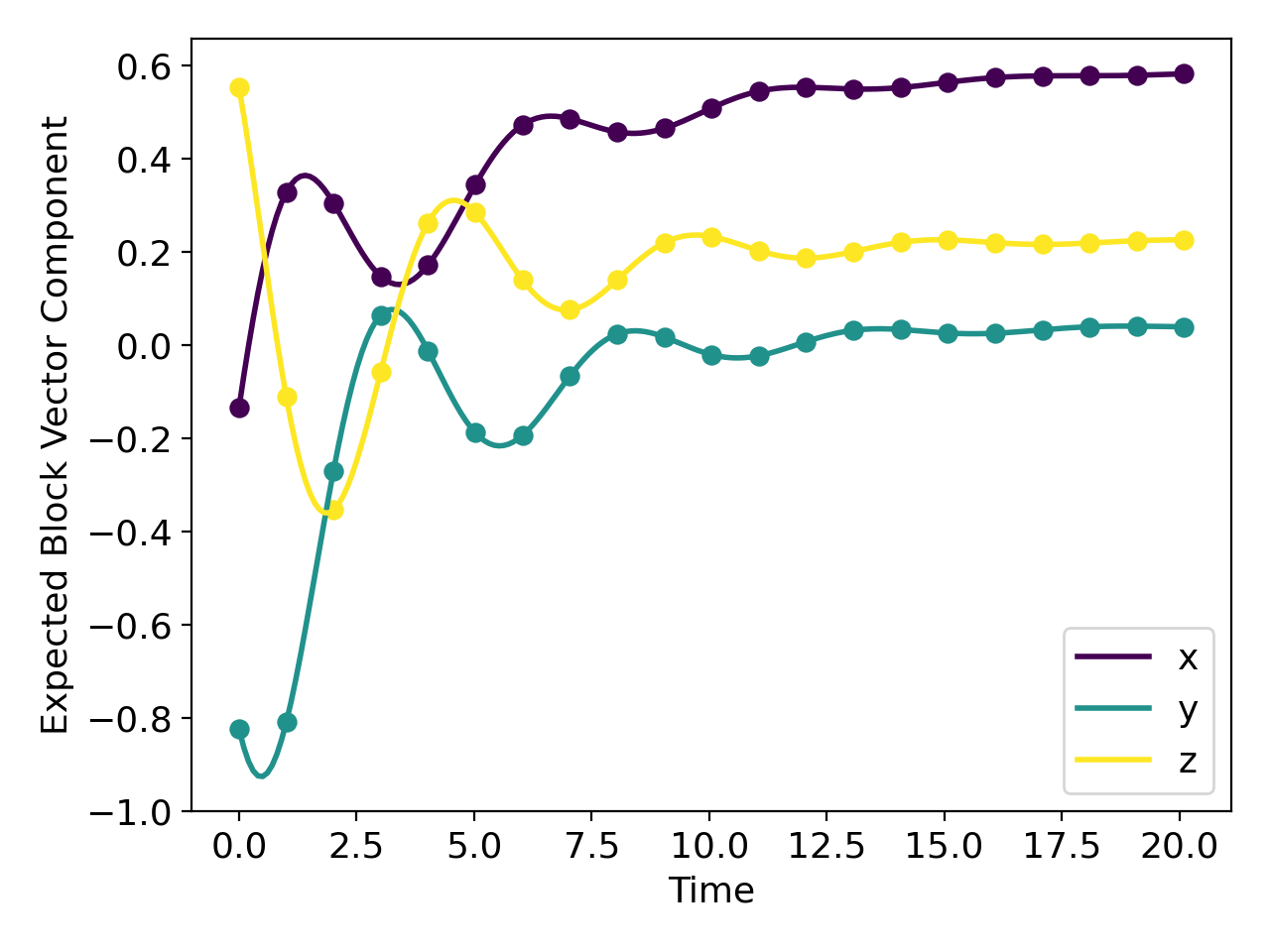}
	\caption{(Color on line) Comparison between the exact time-dependent trajectories (continuous line) and those 
	predicted by the machine-learned regression (points). Here we plot the time-dependent expectation values of the 
	three Pauli operators over the evolution determined by the Lindblad equation, Eq.~(\ref{Limb1Q}). The predictions 
	agree with the exact values up to a floating point error. }
	\label{fig:linblaad}
\end{figure}

As a warm-up example we consider a single qubit evolving via the Lindblad master equation \cite{breuer2002theory}
\begin{equation}\label{Limb1Q}
	\frac{d\hat{\rho}}{dt} = -i[\hat{H}, \hat{\rho}] + 2\hat{L}\hat{\rho}\hat{L}^\dagger - \{\hat{L}^\dagger\hat{L},\hat{\rho}\} ,
\end{equation}
where we have chosen $\hat{H}=0.5\hat{\sigma}_x + 0.3\hat{\sigma}_y + 0.2\hat{\sigma}_z$ and $\hat{L}=\sqrt{0.1}\hat{\sigma}_x$. 
Note that this is non-unitary but Markovian. Here the propagator is of the form $\textbf{x}_{n+1}=A\textbf{x}_n+\textbf{a}$, 
where $A$ is a matrix and $\textbf{a}$ is a vector. However, by defining $\tilde{\textbf{x}}=(1, x_1, x_2,...)$ we can write it in the form 
$\tilde{\textbf{x}}_{n+1}=P\tilde{\textbf{x}}_n$ and thus apply Eq.~(\ref{lin_reg}). Figure \ref{fig:linblaad} compares the 
exacts Bloch vector trajectories with those generated by the learnt propagator. Here four samples were required to learn 
the propagator and the two trajectories agree up to floating point error. 

\section{Predicting non-markovian dynamics}
\label{sec:nMK}
\begin{figure}[h]
	\centering
	\includegraphics[width=0.9\columnwidth]{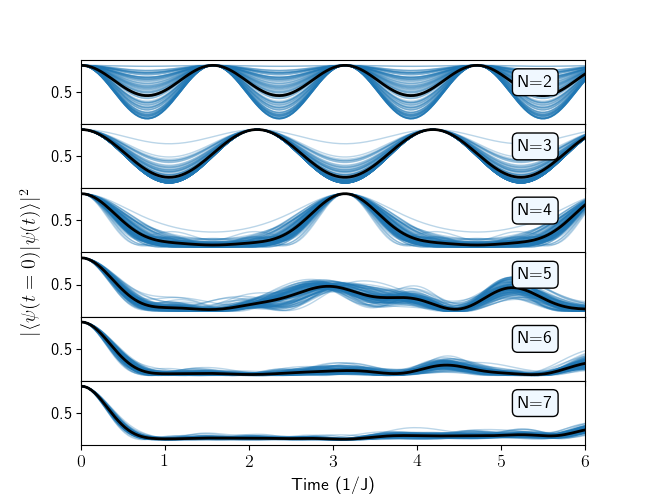}
	\caption{(Color on line) The fidelity of the wavefunction, $|\psi\rangle$, with respect to the initial state, 
	$|\psi(t=0)\rangle$, is computed as a function of time for states initialized randomly. Here, the fidelity is 
	defined as $|\langle\psi(t=0)|\psi(t)\rangle|^2$. The blue lines are the individual trajectories, while the black 
	lines their mean. Note that for $N=2, 3, 4$ the fidelity is periodic in time. The time evolution are for the 
	Heisenberg Hamiltonian of Eq.~(\ref{eq:H}).}
	\label{fig:correlation_time}
\end{figure}

Let us now move to the non-Markovian dynamics. In this case our Hamiltonian is the Heisenberg model of 
Eq.~\eqref{eq:H}, calculated for a number of qubits going from $N=2$ to $N=7$. Information about the time 
evolution can be extracted by plotting the wave-function fidelity, $|\langle\psi(t=0)|\psi(t)\rangle|^2$, as a function 
of time for a range of random initial states, as presented in Figure~\ref{fig:correlation_time}. We note that for 
$N\le4$ the fidelity appears periodic in time over the time interval considered, with periods $T_2=\pi/2$, 
$T_3=2\pi/3$ and $T_4=\pi$, respectively for $N=2, 3$ and 4 (all times are in units of $1/J$). We also note that 
after a time of approximately $1$ ($1/J$) the fidelity approaches zero regardless of $N$. Such time can be 
defined as the characteristic de-correlation time.
Next, we describe our results obtained for the non-Markovian dynamics of single particle Bloch vectors. 

\subsection{Memory}

\begin{figure}[h]
	\centering
	\includegraphics[width=9.5cm]{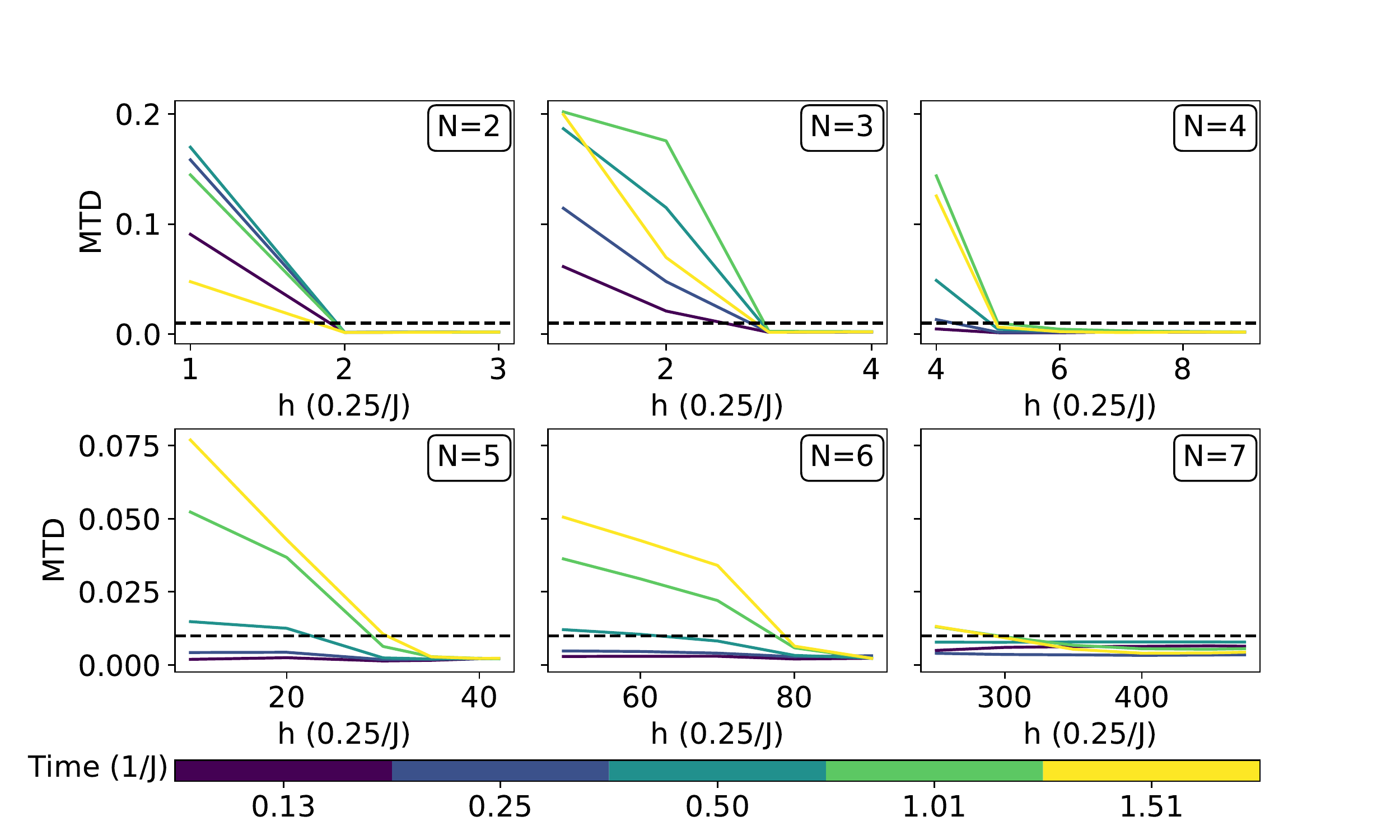}
	\caption{(Color on line) Error of the NN (the mean trace distance - MTD) as a function of the memory, $h$, 
	(in units of $\Delta=0.08\pi\approx0.25/J$), for NNs that propagate at different points in the future (color code). 
	Here the results are for NNs that use as feature the reduced density matrix of all qubits (`all' case).}
	\label{fig:future_a}
\end{figure}

\begin{figure}[h]
	\centering
	\includegraphics[width=9.5cm]{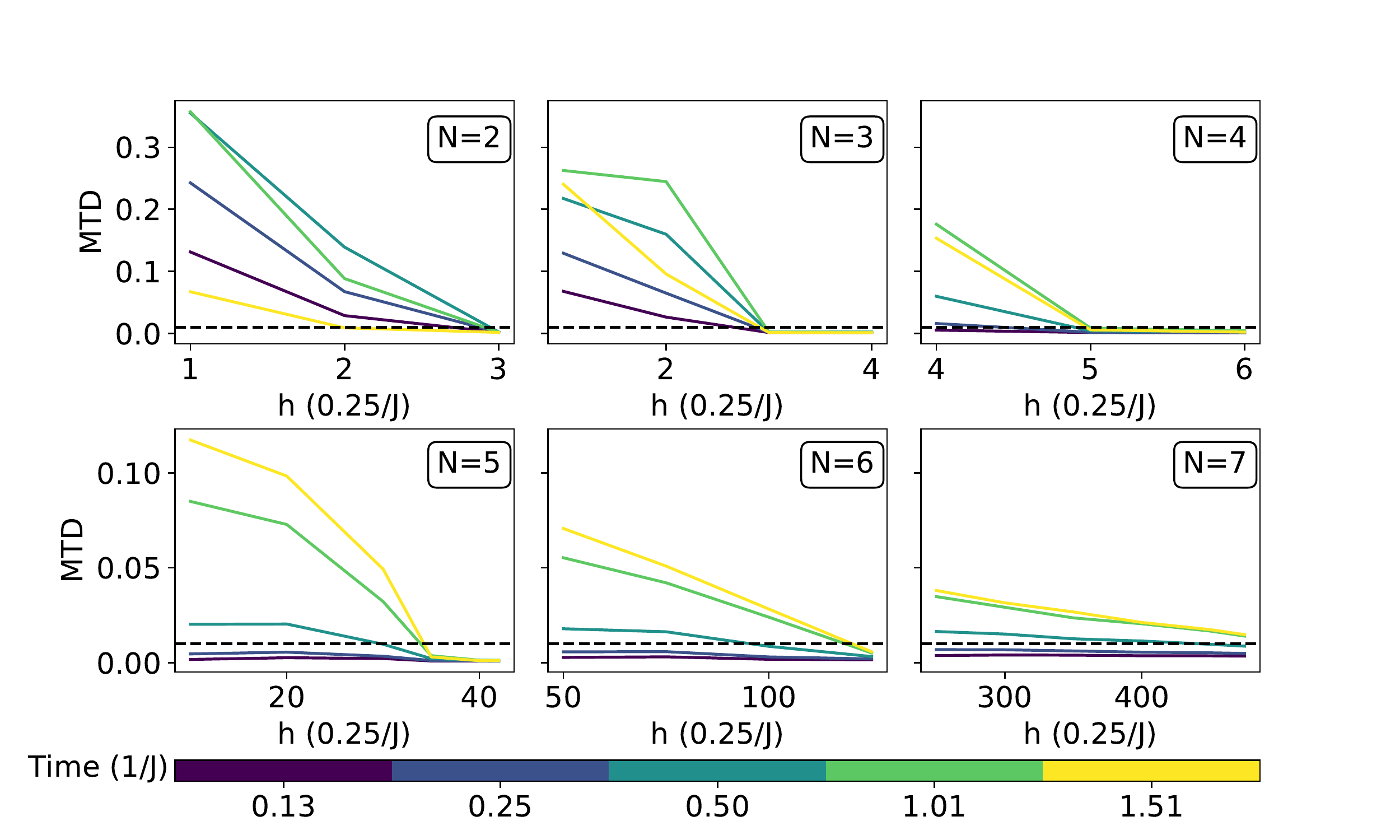}
	\caption{(Color on line) Error of the NN (the mean trace distance - MTD) as a function of the memory, $h$, (in units 
	of $\Delta=0.08\pi\approx0.25/J$), for NNs that propagate at different points in the future (color code). Here the 
	results are for NNs that use as feature the reduced density matrix of a single qubit (`single' case).}
	\label{fig:future_s}
\end{figure}

Let us now consider the dynamics of the single-particle reduced density matrix and evaluate the memory 
needed to propagate a system initialized in a random state. In particular, we set the time step at 
$\Delta=0.08\pi\approx 0.25$ and predict several times into the future. We consider two different cases. 
In the first, figure~\ref{fig:future_a}, the feature vector for the NN comprises the Bloch vectors of all qubits, 
namely we simultaneously follow the dynamics of the single-particle reduced density matrices of all the qubits. 
In the second case, figure \ref{fig:future_s}, the NN propagates the Bloch vector of a single qubit only (note that
the qubit choice here is irrelevant since the system is a ring, namely it has translational symmetry). In the figures 
we plot the NN error, the mean trace distance (MTD), as a function of the memory (in units of the time step, 
$\Delta=0.08\pi$) for NNs that propagate at different times into the future (color code).

In general, in both situations one can always find a sufficiently long memory to converge the NNs to small 
errors (in the figures an error of 0.01 is indicted as a dashed black line). For the `all' case such memory seems
to be rather independent of the final time of propagation, namely it takes approximately the same memory to 
propagate the entire Bloch-vector manifold to either short or long times. This is more evident when the total 
number of qubits is small, while some scattering in the data appears for $N\ge5$. In this case propagating at 
longer times in the future seems to require a deeper memory. Similar results are found when constructing NNs 
using a single Bloch vector as feature (see Fig.~\ref{fig:future_s}), in particular when the qubit count remains 
low ($N\le4$). Most importantly, in both situations the memory needed to converge the NNs increases drastically 
with the number of qubits.

\begin{figure}[h]
	\centering
	\includegraphics[width=1.0\columnwidth]{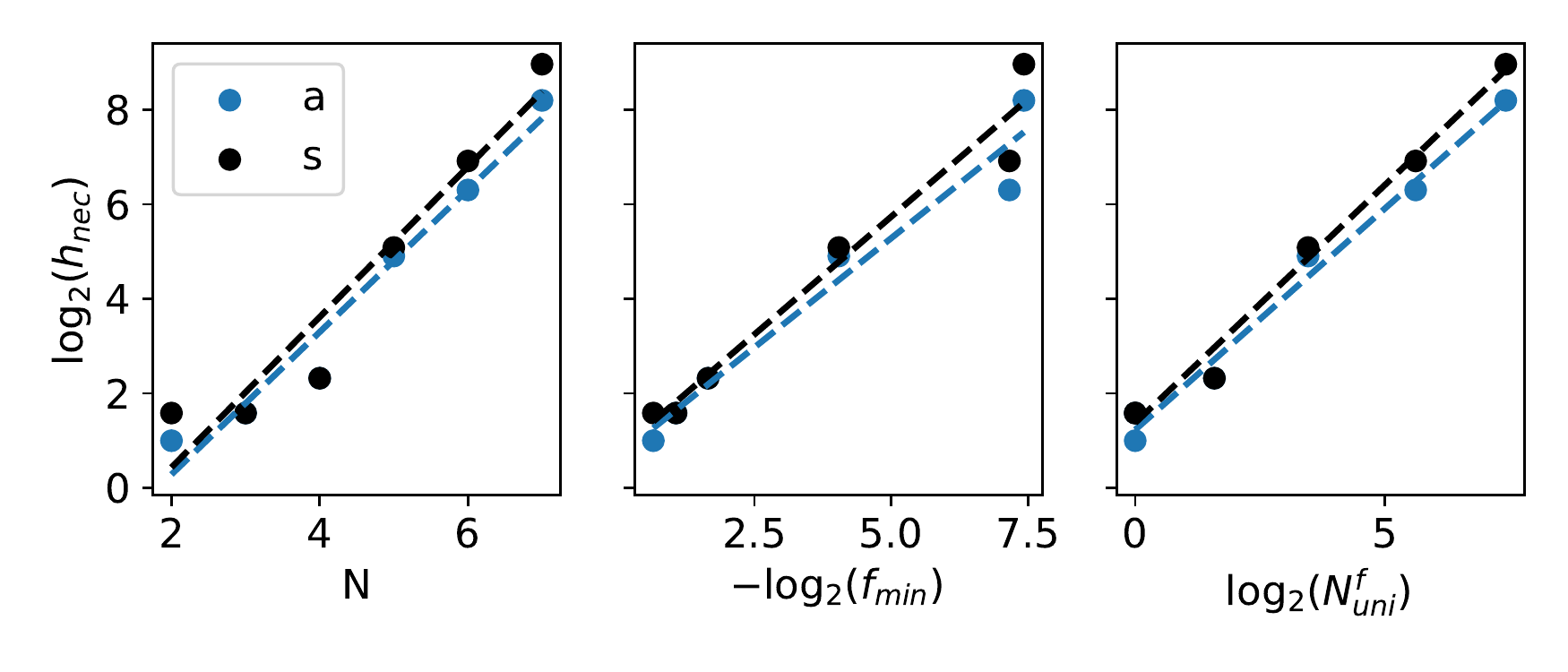}
	\caption{(Color on line) Scaling of the memory needed to achieve accurate propagation (see definition 
	in the text), $h_\mathrm{nec}$, as a function of: (left) the number of qubits, (middle) the minimum frequency, 
	$f_\mathrm{min}$, and (right) the number of unique frequencies of the spectrum of the corresponding 
	Heisenberg chain, $N_\mathrm{uni}^f$. Note that all quantities are plot on a $\log_2$ scale. Light
	(dark) blue symbols are for the `all' (`single') case. Here data are presented for NNs predicting 
	$\Delta=0.48\pi\sim1.5/J$ in the future.}
	\label{fig:scaling}
\end{figure}

The scaling of the memory with the number of qubits is presented in the left-hand side panel of 
figure~\ref{fig:scaling} for NNs propagating $0.48\pi\approx 1.5$ in the future ($\Delta=0.08\pi$). 
This is a time significant longer than the de-correlation time observed in Fig.~\ref{fig:correlation_time}. 
We take the operational definition of `necessary' memory, $h_\mathrm{nec}$, as the memory needed 
to reduce the error below 0.01, and this is plotted on a base-2 logarithmic scale against $N$. We find 
that $h_\mathrm{nec}$ scales exponentially with $N$, namely $h_\mathrm{nec}\propto2^{\alpha N}$, 
with $\alpha\sim1$ (this cannot be determined with precision from our limited number of data points). 
Interestingly, we find little difference in the $h_\mathrm{nec}$ scaling between the `all' and `single' case, 
although for large $N$'s $h_\mathrm{nec}$ is systematically lower when all Bloch vectors are used in 
the NNs.

In Figure \ref{fig:scaling} we also plot $\log_2 (h_\mathrm{nec}$), respectively as a function of 
$-\log_2(f_\mathrm{min})$ (middle panel), where $f_\mathrm{min}$ is the lowest frequency of a 
given system, and as a function of the $\log_2$ of the number of unique frequencies associated to 
the spectrum of the corresponding system (right-hand side panel). The frequencies are computed 
as $\Delta\epsilon_{nm}=\epsilon_n-\epsilon_m$, with $\epsilon_n$ being an eigenvalue of the Heisenberg 
Hamiltonian, obtained by exact diagonalization. We find an extremely good correlation between 
$h_\mathrm{nec}$ and the number of unique frequencies, suggesting that the memory needed for an 
accurate propagation of the time propagation may scale more favourably for systems presenting a limited 
number of frequencies (e.g. in the case of high degeneracy). This correlation, however, may still be accidental,
so that it would be interesting to investigate a many-body Hamiltonian where the multiplicity of the frequency 
spectrum presents a scaling, as a function of the number of particles, different from that of the Hilbert space.

\subsection{Propagation}
% Autoregression
%
In order to show that our NNs can act as true time propagators we need to demonstrate that their time evolution can be
concatenated, namely that one can use the predictions made for time $t$ as an input for the subsequent prediction at time 
$t+\Delta t$. Repeating such a process, an operation called autoregression, allows one to follow the dynamics at, in principle, 
arbitrary times. Such exercise is performed here for the case of six qubits, $N=6$, a propagator with a time time step of 
$\Delta=0.16\pi\sim 0.5/J$ and a memory of $h=45$. We consider NNs using the entire set of Bloch vectors as feature 
(the `all' case) and we initiate the dynamics from a random state. 

\begin{figure}[h]
	\centering
	\includegraphics[width=1.0\columnwidth]{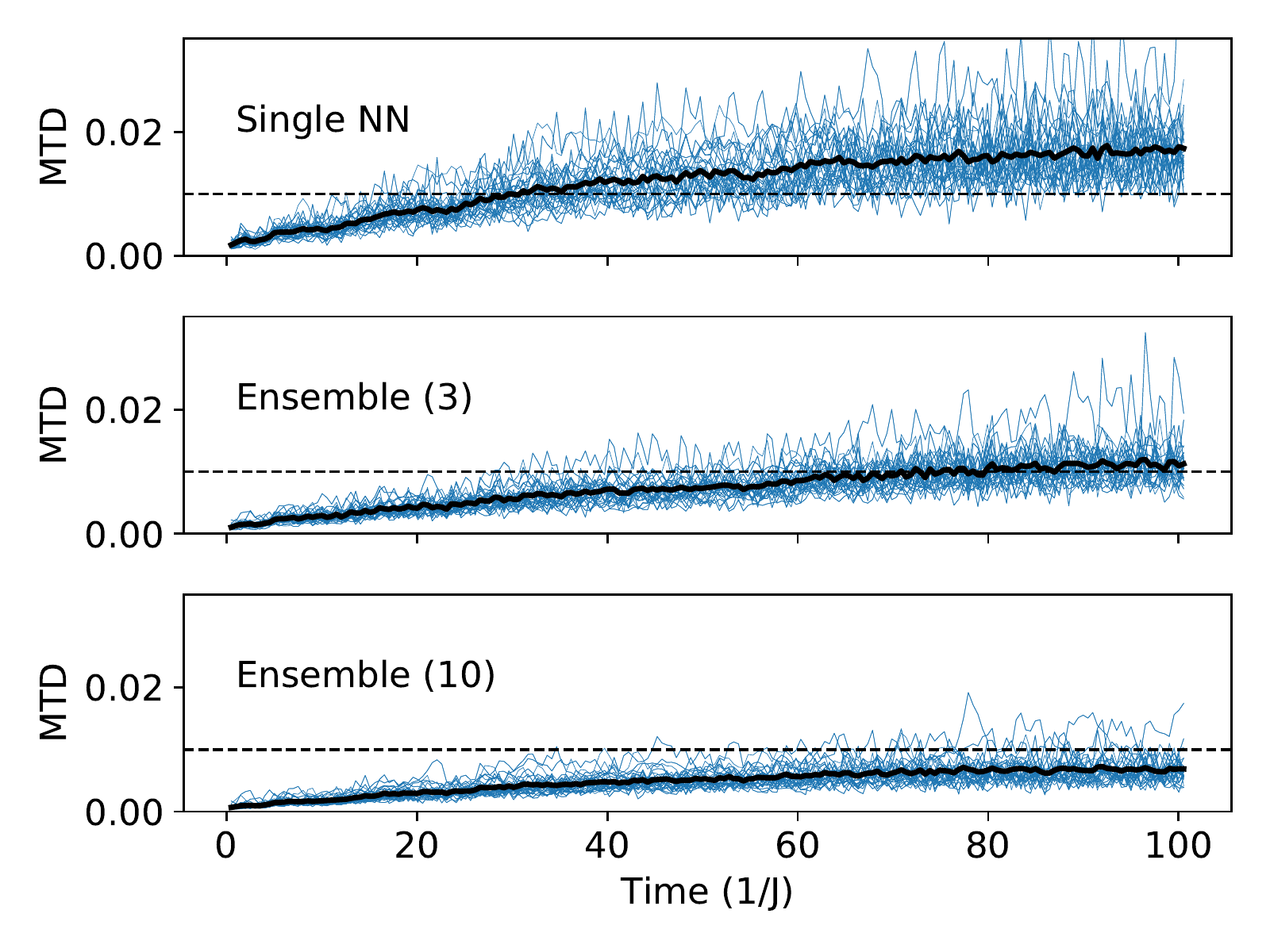}
	\caption{(Color on line) Demonstration of autoregression, where the output of a NN at time $t$ is used as the input
	for the propagation at time $t+\Delta t$. Results are presented for NNs with a time time step of $\Delta=0.16\pi\sim 0.5/J$ 
	and a memory of $h=45$. The system comprises $N=6$ qubits, the initial state is a random state and the NNs use the 
	entire manifold of Bloch vectors as feature. The graphs show the mean trace distance error as a function of time for
	different trajectories (blue lines). The bold line corresponds to the time-averaged mean error. In the top panel we use 
	a single NN, in the middle panel an ensemble of three NNs and in the bottom panel an ensemble of ten. Note that both 
	the time-averaged error and the fluctuations around the average improve as one uses a large number of NNs.}
	\label{fig:prop_error}
\end{figure}

The results of this test are presented in Figure \ref{fig:prop_error}, where we plot the error (the mean trace distance) as a 
function of time for propagation up to $100/J$, namely for about 200 time steps. In order to minimize the error we train 
multiple NNs along the same time trajectories and propagate the Bloch vectors by using the average of the networks' 
predictions. Thus, in Fig.~\ref{fig:prop_error} we show results for a single NN (top panel), an average of three NNs (middle 
panel) and one of ten (bottom panel). Two main conclusions can be drawn from the figure. On the one hand, it is clear that
our NNs are well capable of performing autoregression, with the error growing relatively slowly in time. On the other hand, 
the figure clearly shows that averaging over several NNs significantly improves the predictions; the average is more
accurate and the fluctuations around the average are reduced. This effectively demonstrates that our NNs can be used 
as universal time-evolution operators.

\begin{figure}
	\begin{subfigure}
	\centering
		\includegraphics[width=\linewidth]{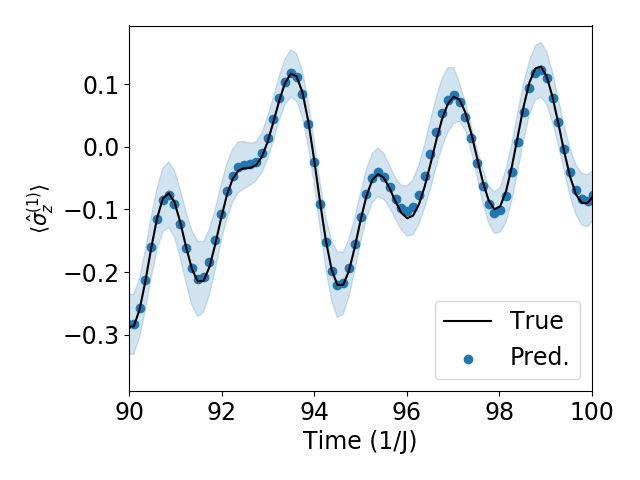}
	\end{subfigure}
	\begin{subfigure}
	\centering
		\includegraphics[width=\linewidth]{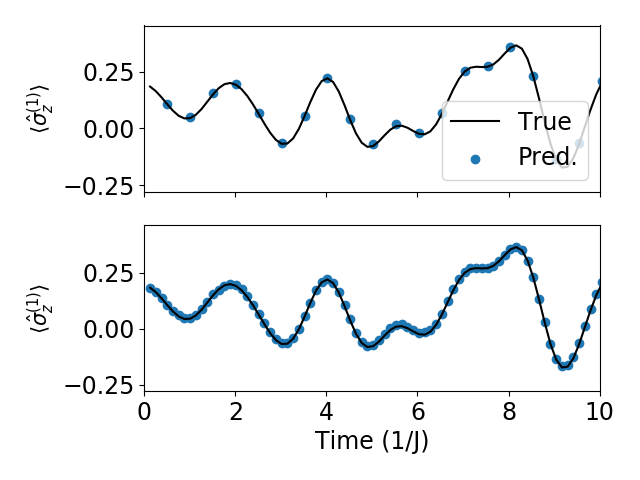}
	\end{subfigure}
	\caption{(Color on line) Time trajectories generated via NN autoregression. Here we consider $N=6$ qubits 
	and monitor the time dependence of the expectation value of $\sigma_z$ of one qubit. The memory in this 
	case is $h=45$. In the top panel we show the predicted trajectory (dots) against the exact one (solid black line), 
	together with the variance of the prediction over ten NNs (blue shadow). The middle panel show the predictions 
	obtained with a time-propagator NN of step $\Delta=0.16\pi$, while the bottom one uses the same predictions 
	as starting point for time propagators with time steps 0.04$\pi$, 0.08$\pi$ and 0.12$\pi$. Note NNs trained
	to predict the dynamics at different time scales can be effectively concatenated.}
	\label{fig:prop}
\end{figure}

When considering an ensemble of NNs one can then use the disagreement among the NNs as a measure of 
the confidence over the prediction. This is essentially the variance of the predicted quantity over the different NNs. 
Such variance is plotted in the top panel of Figure~\ref{fig:prop} for the expectation value of $\sigma_z$ along the 
time trajectory of a single qubit in a system of $N=6$ qubits. Also in this case we simulate the time-evolution of 
$N=6$ qubits with a memory $h=45$. In this particular case the variance does not significantly change over time, 
indicating that the accuracy is largely preserved over the trajectory. However, one can also note that the variance 
is larger along particular branches of the trajectory, where the predictions of the NNs agree less well with the exact 
results. As such, the variance can be used as a measure of the accuracy of the autoregression. Specifically, one can 
monitor the increase in variance and use it as an indicator of the fidelity of the prediction as a function of time.

Finally, we show that NNs trained to propagate the dynamics at different time steps can be concatenated, namely 
that the output of a network of time step $\Delta t_1$ can be used as input for a NN with time step $\Delta t_2$. 
This is shown in the two lower panels of Figure~\ref{fig:prop}. Firstly, we present the trajectory of the expectation 
value of $\sigma_z$ of one qubit, computed with a NN using a time step of $\Delta=0.16\pi$. Then, we use the 
output of such NN as input in NNs of steps $0.04\pi$, $0.08\pi$ and $0.12\pi$. This effectively allows us to 
increase the density of the predicted points along the trajectory. The figure shows clearly that such an operation is 
possible, without any significant loss of accuracy. This is an important results, as it allows one to construct a 
range of NNs, predicting at different times in the future, and use appropriate combinations of them to reach any 
point in time along the trajectory. Crucially, this means that the NNs autoregression can be used to generate 
long-time and arbitrary dense time trajectories.

\subsection{Dynamics initiated from a product state}
\begin{figure}
	\centering
	\includegraphics[width=\columnwidth]{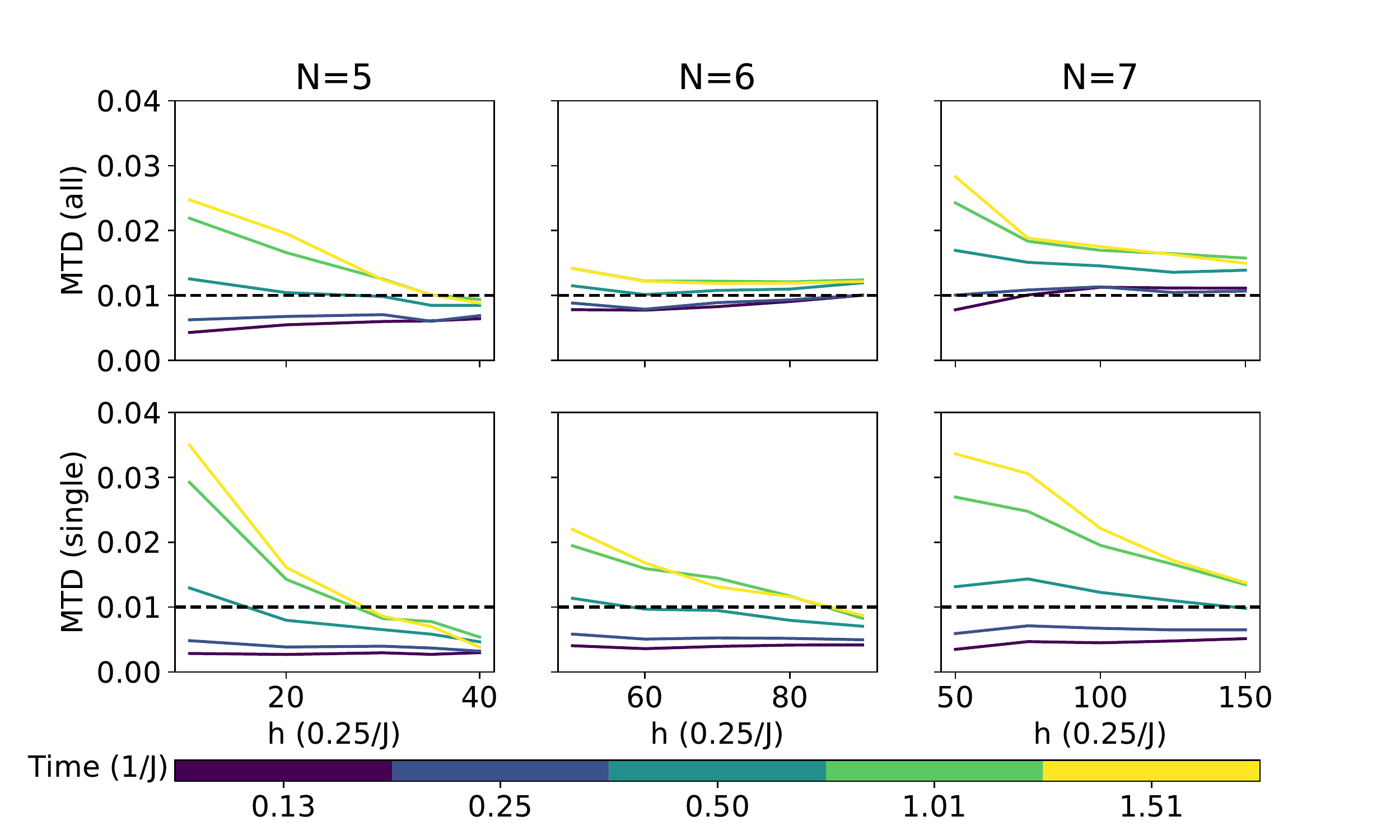}
	\caption{(Color on line) Error of the NN (the mean trace distance) as a function of the memory, $h$, (in units of 
	$\Delta=0.08\pi\approx0.25/J$), for NNs that propagate at different points in the future (color code). Here the 
	results are for NNs that use as feature either the reduced density matrix of all qubits (`all' case - top panels) or of
	a single qubit (`single' case - lower panels).}
	\label{fig:f8futuremps}
\end{figure}

% Product State
%
So far we have analyzed the dynamics initialized from a random state, namely from a state with a high 
entanglement entropy. Here we repeat the analysis for an initial state corresponding to a product state. 
Recalling figure~\ref{fig:sgrowth}, the single-particle entropy of a product state grows with the time, meaning 
that a NN trained from the early-time trajectory will not contain enough information to reproduce the correct 
dynamics and cannot be used in an autoregression. This is because the states encountered during the early 
dynamics of a product state are qualitatively different from those encountered over long times. As such, we 
have now trained NNs by using as an initial state the one evolved from a product state at time 
$0.32\pi\approx 1/J$ (this time is sufficient to reach equilibrium, see Fig.~\ref{fig:sgrowth}), and our results are 
summarised in Figure~\ref{fig:f8futuremps} for system with $N\le5$.

In the figure we show the NN error (the mean trace distance) as a function of the memory for different time 
propagators computed for both the `all' and `single' case. The results are qualitatively similar to those encountered 
for networks trained over dynamics initiated with random states (see Fig.~\ref{fig:future_a} and Fig.~\ref{fig:future_s}), 
except that the memory required to converge the propagator is now, in general, significantly shorter. This follows from 
the fact that the dynamics initiated from a product state spans only a subset of the entire Hilbert space, an observation 
corroborated by the result that product states never evolve to configurations with single-particle entropy close to that 
of random states (see Fig.~\ref{fig:sgrowth}). Such behaviour is not surprising, given the high-symmetry form
of the Hamiltonian.

\begin{figure}
	\centering
	\includegraphics[width=0.9\columnwidth]{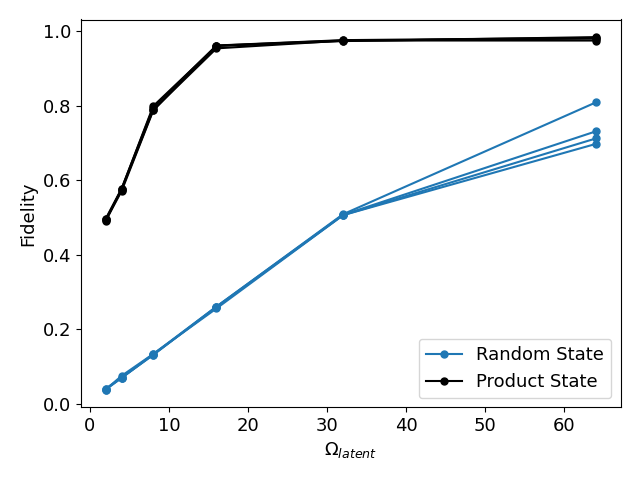}
	\caption{(Color on line) Wavefunction fidelity as a function of the dimension of the latent space (reduced dimension)
	for wavefunction evolved to times 0, 0.25, 0.5 and 1 (in units of $1/J$). Here we compare evolutions initiated from either
	a random or a product state for $N=5$.}
	\label{fig:f11auto}
\end{figure}

We investigate further this aspect by constructing an autoencoder \cite{goodfellow2016deep} performing a 
non-linear compression of the wavefunctions corresponding to both random and product states. To perform this 
task the wavefunction is expanded over the full Fock space $\{|i\rangle\}$ and the real and imaginary coefficients 
of expansion, $\psi_i=\langle i|\psi\rangle$, are taken to form a 2$\Omega$-dimensional vector $\textbf{x}$ with
\begin{equation}
x_j = \begin{cases}
\text{Re}(\psi_{j}) \:\:\:\:\:\quad j \le \Omega\\
\text{Im}(\psi_{j-\Omega}) \quad j > \Omega \\
\end{cases}
\end{equation}
where $\sum_{i=1}^\Omega|\psi_i|^2 = |\textbf{x}|^2 =1$ and $\Omega$ is the dimension of the Hilbert space. 
The encoder and decoder forming the autoencoder are two fully connected NNs, with two 64-nodes hidden layers 
and the ELU activation function. By minimizing the reconstruction error, as measured by the Euclidean distance 
between the vector representations, we can quantify the level of compression that a wavefunction can undergo 
at different times. Figure~\ref{fig:f11auto} shows our results obtained with a dataset of 2,000 samples, with 1,000 
in the training set and 500 in the validation and test ones. For this numerical experiment we consider $N=5$. Clearly, 
a product state can be compressed much more efficiently than a random one (a 15-dimensional latent space appears 
to be sufficient at all times), regardless of the time at which the wavefunction is measured. Note here that for $N=5$ 
the dimension of the Hilbert space is 32. Therefore, when the latent space is 64-dimensional there is no compression 
and the autoencoder just learns the identify function. As such, one expects the fidelity to approach unity regardless 
of the state. The deviation observed for the random state is then attributed to relatively small training set.

\subsection{Frequency of the memory sampling}

\begin{figure}
	\centering
	\includegraphics[width=1.0\columnwidth]{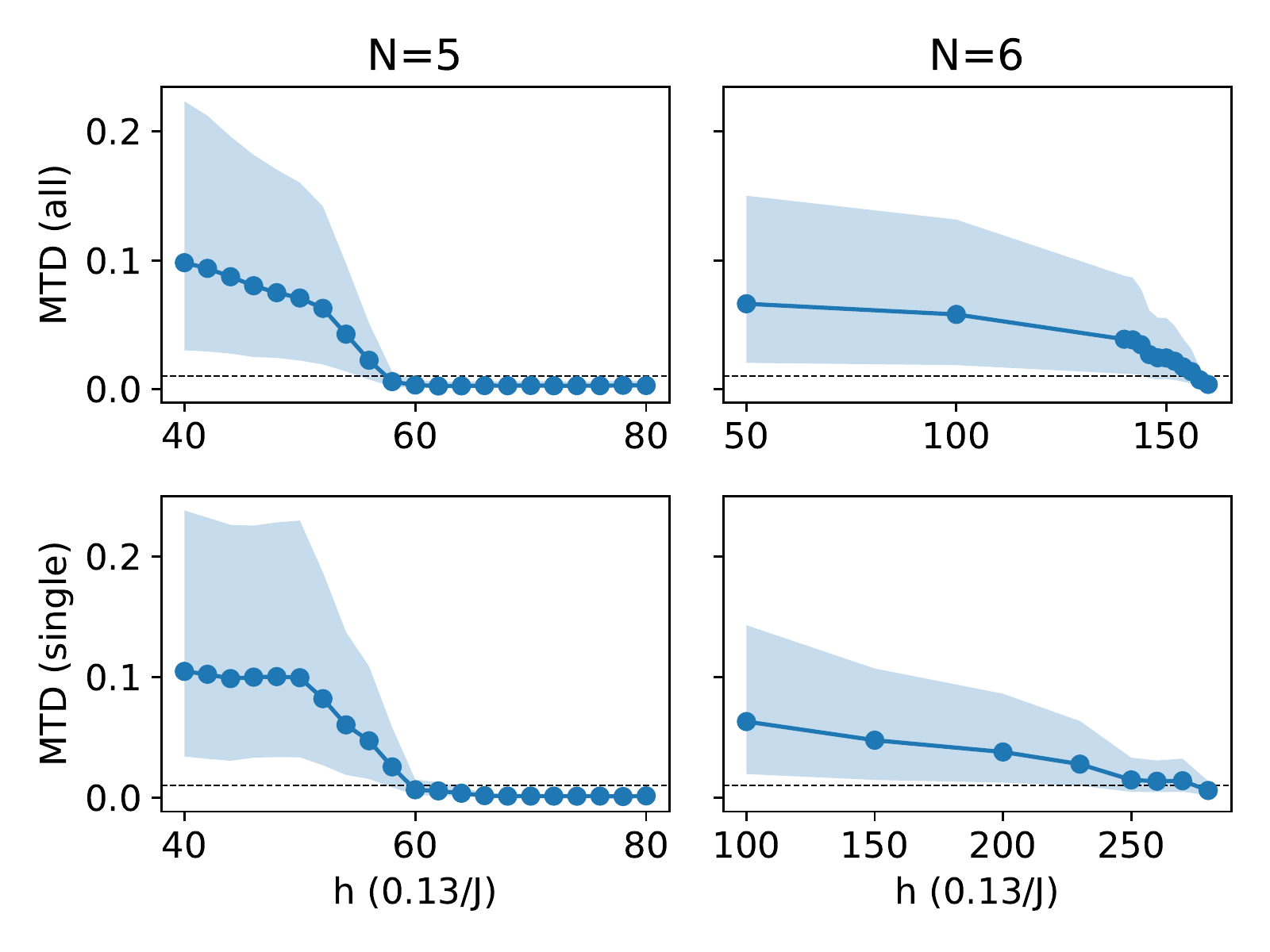}
	\caption{(Color on line) Error of the NN (the mean trace distance) as a function of the memory, $h$, (in units of 
	$\Delta=0.04\pi\sim0.13/J$), for $N=5$ and $N=6$ and both the `all' (upper panels) and `single' (lower panels) case.
	In this case the NN propagates $0.48\pi\approx 1.5/J$ in the future. The shaded region corresponds to the 25th and 
	75th error percentiles.}
	\label{fig:f10fine}
\end{figure}

We now investigate further the nature of the memory required for non-Markovian 
dynamics. The first question we want to address is whether the convergence of the memory is smooth or 
sharp, namely whether the error in the propagation drops sharply when one reaches the required memory. 
This is explored in Figure~\ref{fig:f10fine}, where we present the error of the NN (the mean trace distance 
of a single NN for all the samples in the training set) as a function of the memory duration for propagation 
to a time of $0.48\pi\approx 1.5/J$, where the memory is sampled with the fine time resolution of 
$0.04\pi\approx 0.13/J$. The exercise is performed for both $N=5$ (left-hand side panels) and $N=6$ (right-hand 
side panels), in the `all' (top panels) and `single' (bottom panels) cases. The graphs clearly show a very sharp 
transition, with the error abruptly reducing below the 0.01 threshold as soon as the memory reaches a critical 
duration. This means that, if the memory is accurately sampled, one will just need to construct a NN time 
propagator with a memory longer than the critical one. In fact, considering longer memories does not improve 
the convergence. In other words, it appears that the range of the time-propagator kernel is finite and well defined 
for any given system.

\begin{figure}
	\centering
	\includegraphics[width=1.0\columnwidth]{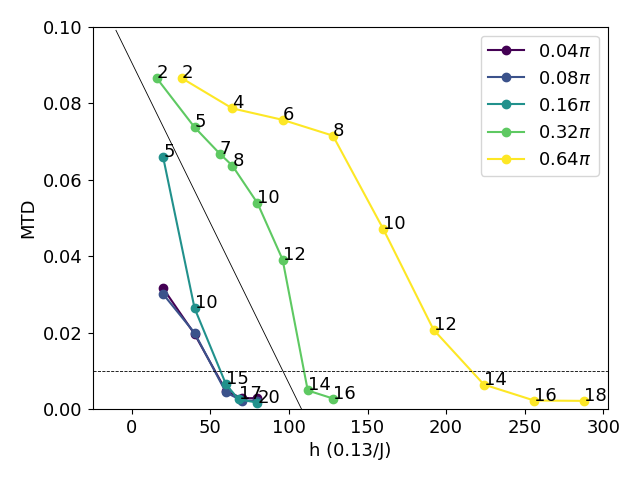}
	\caption{(Color on line) NN error (the mean trace distance) as a function of the memory for the propagation to the 
	time $0.48\pi\approx 1.5/J$ of a $N=5$ system. Different curves correspond to memories sampled with different time 
	steps. The numbers on each curve correspond to the number of time steps included in the memory (the dimension of the
	time component of the feature vector), while the $x$-axis scale is absolute ($\Delta=0.04\pi$). The black diagonal line 
	separates the results for which the memory duration is independent from the sampling.}
	\label{fig:f9course}
\end{figure}

Having established that there is a critical memory for each system, one now needs to find out how finely such 
memory should be sampled. This question is answered in Figure~\ref{fig:f9course}. Here we have constructed 
a single NN to propagate a $N=5$ system $0.48\pi\approx1.5/J$ in the future (`all' case). The memory is then 
sampled with different time steps, going from 0.04$\pi$ to 0.64$\pi$. Surprisingly, we find that when the memory 
is finely sampled ($\Delta=0.04\pi$, $0.08\pi$ and $0.16\pi$) its duration does not change, namely feature
vectors of different length can equally well predict the dynamics as long as enough time-evolution history is learned. 
In contrast, more coarse samplings ($\Delta>0.16\pi$) require longer memories. Intriguingly, in this case of coarse 
sampling, one approximately needs the same number of time steps (about 14), although of different duration, to 
converge the NN.

Notably, the highest frequency found in the spectrum of the $N=5$ spin chain is $f_\mathrm{max}=0.99J$. 
According to the Shannon-Nyquist sampling theorem \cite{allen2004signal} the largest possible period required to 
sample a time-dependent dynamics is $\Delta_\mathrm{max}=1/2f_\mathrm{max}$, which in our case is 
$\Delta_\mathrm{max}\approx0.16\pi$. Intriguingly, this corresponds to the critical sampling time above which the 
memory is no longer time-step independent. We can then conclude that the two regimes found in Figure~\ref{fig:f9course} 
are simply separated by the Shannon-Nyquist limit. In general, for a linear model the dynamics cannot be reproduced 
at all if sampled at a time step larger than $\Delta_\mathrm{max}$. Here, however, our time-propagator (the NN) is 
highly non-linear and the Shannon-Nyquist limit can be avoided by sampling longer memories at a coarser resolution. 
Further tests using non-linear memory samplings have not given conclusive results.

\subsection{Information locality}

\begin{figure}
	\centering
	\includegraphics[width=0.9\columnwidth]{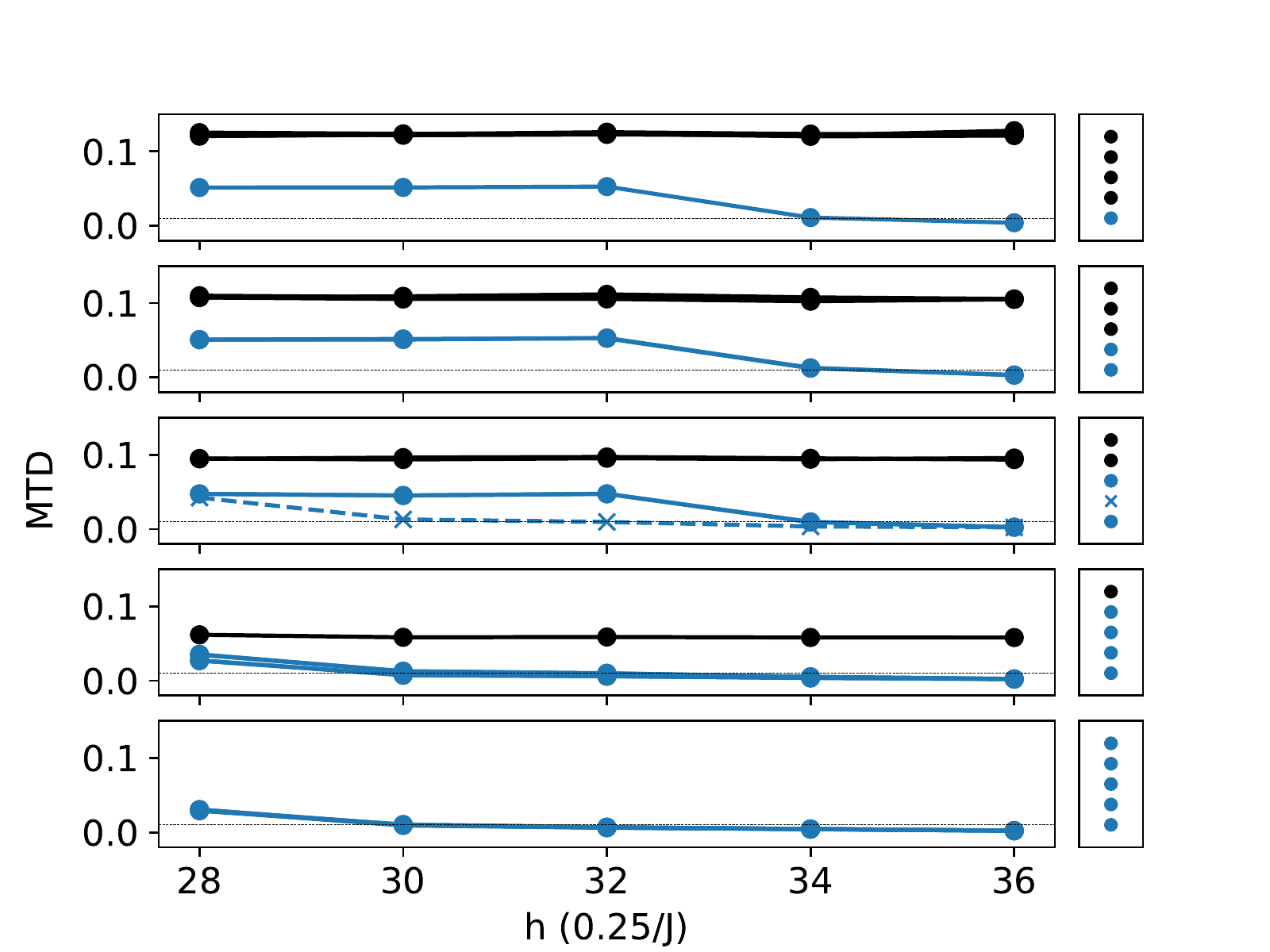}
	\caption{(Color on line) NN error (the mean trace distance) as a function of the memory for the propagation to the 
	time $0.48\pi\approx 1.5/J$ of a $N=5$ system. Here the NN is trained by including only the Bloch vector of
	$m$ qubits, while the remaining $N-m$ are not considered. In the graphs the error over the qubits included in the
	training is in blue, while that of the qubits outside the NN feature vector are in black.}
	\label{fig:local1}
\end{figure}

Finally we investigate how the information included in the feature vector affects our ability to make predictions. For this 
experiment we take the $N=5$ spin chain with dynamics initiated at a random initial state. In this case we compute
the error (the mean trace distance) of a single NN as a function of the memory for a generic case where the feature 
vector contains the Bloch vectors of $m$ qubits. The error is then computed over both the $m$ qubits used in the 
feature vector and the remaining $N-m$ ones. With this notation $m=1$ and $m=5$ corresponds to the `single' and `all' 
case, respectively. Our results are summarized in Figure~\ref{fig:local1}.

In general we find that the models cannot satisfactory predict the evolution of qubits, whose Bloch vector was not 
included in the training, although the error reduces as $m$ gets larger. Interestingly, this is true even for the $m=4$ 
case, where only the Bloch vector of one qubit is left outside the NN feature vector. A perhaps more intriguing feature 
is found for the $m=3$ case. In this situation the three qubits used for the training are inequivalent, since only one neighbour 
other qubits included in training the model. In this case we found that the memory required to bring the error below threshold 
is different depending on the location of the qubit. This suggests that there may be a trade off between time and space 
locality in the time propagator kernel. Similar results are also found for the $N=6$ case (not presented here), where several 
inequivalent configurations can be designed. 

\section{Conclusion}
In this work we have explored the use of machine learning to propagate quantum systems in time, using the
Heisenberg Hamiltonian as many-body model. For Markovian dynamics the time propagator can be learnt easily 
with a linear regression, as long as the training dataset is sufficiently large. In contrast, the propagation of 
non-Markovian systems requires a history, meaning that the system state at a given time is determined by a 
number of states in the past. The duration of such memory appears to scale exponentially with the system size, 
regardless of whether one uses the density matrix of a single qubit or of the entire ensemble as feature. However, shorter
memories are required when the system explores only a sub-set of the available spectrum during the time evolution,
as in the case of dynamics initiated from a product state.

Crucially, we have shown that the machine-learning propagators can be concatenated in an autoregression, meaning
that the state evolved with one neural network can be used as input for another propagation. This allows us to propagate
at arbitrary long times with any desired resolution. Our method can then be applied to quantum dynamical data generated 
from any computational scheme, whether it be propagation, tensor networks~\cite{PAECKEL2019167998}, restricted Boltzmann 
machines~\cite{carleo2017solving}, or data obtained experimentally. Furthermore, by using an ensemble of machine-learning
propagators we can maintain accuracy for a large number of iterations and constantly estimate the error of our
predictions.

Finally, we have investigate in detail the time resolution needed to represent the system memory, and found two regimes
separated by the Shannon-Nyquist limit. Namely, when the memory is sampled at a time step shorter than the period
corresponding to the fastest frequency of the system, the memory remains constant. This means that one has to sample 
a fixed time interval but different time steps can be used. In contrast, if the time step is longer than such period, the 
required memory is no longer constant, but the total number of time steps is. This demonstrates that the non-linearity 
built in the neural networks can overcome the limitation set by the Shannon-Nyquist sampling theorem.

For future studies it will be useful to compare the method presented here with the other approaches proposed for 
solving the reduced dynamics of the system, such as the discussed Markovian embeddings, the Nakajima-Zwanzig 
technique and also the transfer tensor methods. In particular, it will be interesting to see if a direct connection between 
the memory kernel appearing in the Nakajima-Zwanzig equation and the history depth $h$ that we observed for the 
trained NNs can be made. 

\subsection*{Acknowledgment}
This work has been supported by the Irish Research Council Postgraduate Scholarship (J.N.) and Advance Laureate
Award (S.S. - Award IRCLA/2019/127). L.C. and G.K. acknowledge Science Foundation Ireland for financial support 
through Career Development Award 15/CDA/3240. We acknowledge the DJEI/DES/SFI/HEA Irish Centre for High-End 
Computing (ICHEC) and Trinity Centre for High Performance Computing (TCHPC) for the provision of computational 
resources. The authors thank Plamen Stamenov and John Goold for interesting discussion.

\appendix
\section{Nakajima-Zwanzig equation}
\label{app:NZ}

Starting from the Liouville-Von Neumann equation,
\begin{equation}
\label{eq:Liouville}
\partial_t \hat{\rho} = \frac{i}{\hbar} \left[ \hat{\rho},\hat{H} \right] = \hat{L} \hat{\rho}\:,
\end{equation}
one splits the density operator up, $\hat{\rho} = (\hat{\mathcal{P}}+\hat{\mathcal{Q}}) \hat{\rho}$, by means of the projection 
operators $\hat{\mathcal{P}}$ and $\hat{\mathcal{Q}}$, with $\hat{\mathcal{Q}}=\hat{\mathcal{I}}-\hat{\mathcal{P}}$
($\mathcal{I}$ is the identity). The definition of $\hat{\mathcal{P}}$ and $\hat{\mathcal{Q}}$ is somewhat unusual. 
Consider a reference state $\hat{\rho}_\mathrm{B}$ of the environment and then define $\hat{\mathcal{P}}$ as the 
projector such that
\begin{equation}
\hat{\rho}_{\text{red}}= \hat{\mathcal{P}} \hat{\rho} = \text{Tr}_\mathrm{B} ( \hat{\rho} ) \otimes \hat{\rho}_\mathrm{B}\:,
\end{equation}
where $\rho_\mathrm{red}$ is the reduced density matrix of the degrees of freedom of interest. Thus $\hat{\mathcal{P}}$ 
is defined as the partial tracing out of the environment and then the product of the resulting state with the predetermined 
environmental state, which is typically time independent. The definition of $\hat{\mathcal{Q}}$ then follows as
\begin{equation}
\hat{\mathcal{Q}} \hat{\rho} = \hat{\rho} - \text{Tr}_\mathrm{B} ( \hat{\rho} ) \otimes \hat{\rho}_\mathrm{B}\:.
\end{equation}
Although unusual the projectors are perfectly well defined and it is easy to show that
\begin{align}
\hat{\mathcal{P}}^2 =\hat{\mathcal{P}}\:, \\
\hat{\mathcal{Q}}\hat{\mathcal{P}} = \hat{\mathcal{P}}\hat{\mathcal{Q}} = 0\:.
\end{align}

Starting from Eq.~(\ref{eq:Liouville}) one inserts $\mathcal{I}=\mathcal{P}+\mathcal{Q}$ to obtain
   \begin{align}
    & \partial_t  \hat{\mathcal{P}}  \hat{\rho} =  \hat{\mathcal{P}}  \hat{L}  \hat{\mathcal{P}} \hat{\rho} + \hat{\mathcal{P}}  \hat{L} \hat{\mathcal{Q}} \hat{\rho}\:, \\
    & \partial_t  \hat{\mathcal{Q}} \hat{\rho} =  \hat{\mathcal{Q}} \hat{L} \hat{\mathcal{Q}}  \hat{\rho} +  \hat{\mathcal{Q}}    \hat{L} \hat{\mathcal{P}}  \hat{\rho}\:,
  \end{align} 
and the second identity can be formally solved as
\begin{align}
\hat{\mathcal{Q}} \hat{\rho} = e^{\hat{\mathcal{Q}} \hat{L} t} \hat{\mathcal{Q}} \hat{\rho}(0)  + \int_0^t dt' e^{\hat{\mathcal{Q}} \hat{L} t'} \hat{\mathcal{Q}} \hat{L} \hat{\mathcal{P}} \hat{\rho}(t-t')\:.
\end{align}
This can then be inserted back into the first equation to give
\begin{align}
   \partial_t  \hat{\mathcal{P}} \hat{\rho} = \hat{\mathcal{P}}\hat{L} \hat{\mathcal{P}} \hat{\rho} + \hat{\mathcal{P}} \hat{L} e^{\hat{\mathcal{Q}} \hat{L} t} \hat{Q} \hat{\rho}(0) + \\
   \hat{\mathcal{P}} \hat{L}  \int_0^t dt' e^{\hat{\mathcal{Q}} \hat{L} t'} \hat{\mathcal{Q}} \hat{L} \hat{\mathcal{P}} \hat{\rho}(t-t')\:.
\end{align}
Finally, if we assume that at the time $t=0$ the system is a product state $\hat{\rho}(0) = \hat{\rho}_\mathrm{A} \times \hat{\rho}_\mathrm{B}$, 
then $\hat{\mathcal{P}} \hat{\rho}(0) = \hat{\rho} (0) $ and $\hat{\mathcal{Q}} \hat{\rho}(0) = 0 $, so that we can drop the middle term to obtain
\begin{equation}
\label{eq:NZ}
\partial_t [ \hat{\mathcal{P}} \hat{\rho}] =  \hat{\mathcal{P}} \hat{L}  [ \hat{\mathcal{P}} \hat{\rho}] +\int_0^t dt' \hat{\mathcal{K}}(t') [ \hat{\mathcal{P}} \hat{\rho}(t-t')]\:,
\end{equation}
where the time kernel writes
\begin{equation}
\hat{\mathcal{K}} (t) = \hat{\mathcal{P}} \hat{L} e^{\hat{\mathcal{Q}} \hat{L} t} \hat{\mathcal{Q}} \hat{L} \hat{\mathcal{P}}\:.
\end{equation}

\bibliographystyle{unsrt}
%\bibliography{refs}

\begin{thebibliography}{hbp}
\bibitem{NoteD}Throughout this work we use the trace distance, $D$, as our metric for comparing density matrices with 
$D = \frac{1}{2}|\sigma-\rho|$. Note that $0 \le D \le 1$ and for Bloch vectors $\textbf{x}$ and $\textbf{y}$ it reduces to 
$\frac{1}{2}|\textbf{x}-\textbf{y}|$.
\bibitem{Note}Since linear regressions require the weight matrix to be real we must write the wavefunction as a $2\Omega$ 
vector and, hence, require $2\Omega$ samples. For the density matrix one needs to evaluate $\Omega(1)+1/2(\Omega^2-\Omega)(2)=\Omega^2$ 
real numbers and hence we need $\Omega^2$ samples. We subtract 1 from both as we have the normalization constraint.
\end{thebibliography}

\end{document}